\documentclass[aps,prd,preprint,superscriptaddress,tightenlines]{revtex4}
\newcommand{\PRE}[1]{{#1}}   

\usepackage{bm}
\usepackage{epsfig}

\newcommand{\postscript}[2]{\setlength{\epsfxsize}{#2\hsize}
   \centerline{\epsfbox{#1}}}

\newcommand{\mstar}{M_{\ast}}
\newcommand{\md}{M_D}
\newcommand{\mbh}{M_{\text{BH}}}
\newcommand{\mbhmin}{M_{\text{BH}}^{\text{min}}}
\newcommand{\xmin}{x_{\text{min}}}

\newcommand{\mev}{\text{MeV}}
\newcommand{\gev}{\text{GeV}}
\newcommand{\tev}{\text{TeV}}
\newcommand{\pb}{\text{pb}}
\newcommand{\cm}{\text{cm}}
\newcommand{\km}{\text{km}}
\newcommand{\g}{\text{g}}
\newcommand{\s}{\text{s}}
\newcommand{\yr}{\text{yr}}
\newcommand{\sr}{\text{sr}}

\newcommand{\etal}{{\em et al.}}
\newcommand{\ie}{{\em i.e.}}
\newcommand{\eg}{{\em e.g.}}

\newcommand{\eqref}[1]{Eq.~(\ref{#1})}

\begin{document}

\preprint{
\hfil
\begin{minipage}[t]{3in}
\begin{flushright}
\vspace*{.4in}
MIT--CTP--3221\\
UCI--TR--2001--43\\
MADPH--02--1255\\
hep-ph/0202081
\end{flushright}
\end{minipage}
}

\title{
\PRE{\vspace*{1.5in}}
Detecting Microscopic Black Holes with Neutrino Telescopes
\PRE{\vspace*{.3in}}
}

\author{Jaime Alvarez-Mu\~niz}
\affiliation{Bartol Research Institute, University of Delaware,
Newark, DE 19716
\PRE{\vspace*{.1in}}
}
\author{Jonathan L.~Feng}
\affiliation{Center for Theoretical Physics, Massachusetts Institute
of Technology, Cambridge, MA 02139
\PRE{\vspace*{.1in}}
}
\affiliation{Department of Physics and Astronomy, University of
California, Irvine, CA 92697
\PRE{\vspace*{.1in}}
}
\author{Francis Halzen}
\affiliation{Department of Physics, University of Wisconsin, 1150
University Avenue, Madison, WI 53706
\PRE{\vspace*{.3in}}
}
\author{Tao Han}
\affiliation{Department of Physics, University of Wisconsin, 1150
University Avenue, Madison, WI 53706
\PRE{\vspace*{.3in}}
}
\author{Dan Hooper%
\PRE{\vspace*{0.2in}}
}
\affiliation{Department of Physics, University of Wisconsin, 1150
University Avenue, Madison, WI 53706
\PRE{\vspace*{.3in}}
}

\begin{abstract} 
\PRE{\vspace*{.2in}} 
If spacetime has more than four dimensions, ultra-high energy cosmic
rays may create microscopic black holes.  Black holes created by
cosmic neutrinos in the Earth will evaporate, and the resulting
hadronic showers, muons, and taus may be detected in neutrino
telescopes below the Earth's surface.  We simulate such events in
detail and consider black hole cross sections with and without an
exponential suppression factor. We find observable rates in both
cases: for conservative cosmogenic neutrino fluxes, several black hole
events per year are observable at the IceCube detector; for fluxes at
the Waxman-Bahcall bound, tens of events per year are possible.  We
also present zenith angle and energy distributions for all three
channels.  The ability of neutrino telescopes to differentiate
hadrons, muons, and possibly taus, and to measure these distributions
provides a unique opportunity to identify black holes, to
experimentally constrain the form of black hole production cross
sections, and to study Hawking evaporation.
\end{abstract}


\maketitle

\section{Introduction}

The possibility that we live in $D = 4+n > 4$ spacetime dimensions has
profound implications.  In particular, if gravity propagates in these
extra dimensions, the fundamental Planck scale $\md$ at which gravity
becomes comparable in strength to other forces may be far below $M_4
\sim 10^{19}~\gev$, leading to a host of potential signatures for high
energy physics~\cite{Antoniadis:1998ig}.

Among the most striking consequences of low-scale gravity is the
possibility of black hole creation in high-energy particle
collisions~\cite{Amati:1987wq,'tHooft:rb,Argyres:1998qn,Banks:1999gd,%
Emparan:2000rs,Giddings:2000ay,Giddings:2001bu,Dimopoulos:2001hw}.  In
most gravitational processes, such as those involving graviton
emission and exchange, analyses rely on a perturbative description
that breaks down for center-of-mass energies of $\md$ and above.  In
contrast, black hole properties are best understood for energies above
$\md$, where semi-classical and thermodynamic descriptions become
increasingly valid~\cite{Hawking:1975sw}.  In principle, then, black
holes provide a robust probe of extra dimensions and low-scale
gravity, as long as particle collisions with center-of-mass energies
above $\md \sim 1~\tev$ are available.

Nature provides interactions at the necessary energies in the form of
cosmic rays with energies above $10^{10}~\gev$.  In collisions with
nucleons, these cosmic rays probe center-of-mass energies exceeding
100 TeV, beyond both current man-made colliders and those of the
foreseeable future.  Cosmic neutrinos may create black holes deep in
the Earth's atmosphere, resulting in spectacular signals of giant air
showers in ground arrays and fluorescence
detectors~\cite{Feng:2001ib,Anchordoqui:2001ei,Emparan:2001kf,%
Ringwald:2001vk,Anchordoqui:2001cg}.  With a handful of events,
standard model (SM) and most alternative explanations may be
excluded~\cite{Feng:2001ib,Anchordoqui:2001cg} by comparison with
rates for Earth-skimming neutrinos~\cite{Bertou:2001vm,%
Feng:2001ue,Kusenko:2001gj}, and with more events, black holes may be
identified through their shower
characteristics~\cite{Anchordoqui:2001ei}.  Bounds have been
derived~\cite{Ringwald:2001vk,Anchordoqui:2001cg} from the absence of
such showers in current data from Fly's Eye~\cite{Baltrusaitis:mt} and
AGASA~\cite{agasa}. For conservative fluxes and the geometric black
hole cross section, the AGASA data require $M_D \agt 1.3-1.8~\tev$ for
$n\ge 4$, the most stringent constraint to
date~\cite{Anchordoqui:2001cg}.  At the same time, the Auger
Observatory, scheduled for completion by 2004, may observe tens of
black hole events per
year~\cite{Feng:2001ib,Ringwald:2001vk,Anchordoqui:2001cg}.  Related
phenomena related to $p$-brane production may also be observed in
cosmic rays~\cite{Ahn:2002mj,Jain:2002kf}.

Here we examine the possibility of detecting and studying black holes
produced by cosmic neutrinos in neutrino telescopes. Several
large-scale neutrino telescope projects are underway, including
IceCube~\cite{IceCube} in the Antarctic ice, and
ANTARES~\cite{ANTARES} and NESTOR~\cite{NESTOR} in the Mediterranean.
Among the many possible black hole signatures, such detectors are most
sensitive to contained hadronic showers and through-going muons and
taus from the evaporation of black holes produced in the Earth's
crust.  The possibility of black hole detection in neutrino telescopes
has recently been studied in Ref.~\cite{Kowalski:2002gb}; where
possible, we will compare our results to those of
Ref.~\cite{Kowalski:2002gb} below. For a preliminary study, see also
Ref.~\cite{Uehara:2001yk}.

For TeV-scale gravity and conservative flux assumptions, we find that
IceCube could detect several black holes per year.  These rates may be
enhanced by larger fluxes, and observable rates are possible even
given a postulated exponential suppression factor in the black hole
cross section~\cite{Voloshin:2001vs,Voloshin:2001fe}.  The relative
event rates in the three channels may differ from the SM, and the
energy and angle distributions of black hole events are also
distinctive.  These will not only help identify black holes, but may
also constrain parameters such as $n$ and $M_D$, and determine if
suppression factors in the cross section are present or absent. The
search for black holes in neutrino telescopes therefore complements
black hole searches in other cosmic ray detectors, as well as searches
for the effects of perturbative gravity processes at center-of-mass
energies below $\md$~\cite{Nussinov:1999jt,Jain:2000pu,Tyler:2001gt,%
Alvarez-Muniz:2001mk}.

\section{Black Hole Production and Evaporation}
\label{sec:production}

In $D=4+n$ dimensions, gravity is described by the Einstein action
\begin{equation}
S_E = \frac{1}{8 \pi G_D} \int d^{4+n}x \sqrt{-g} \,
\hbox{$1\over 2$} {\cal R} \ ,
\end{equation}
where $G_D$ is the $D$-dimensional Newton's constant.  To define the
fundamental Planck scale $M_D$, we adopt the convention
\begin{equation}
\frac{1}{8\pi G_D} = \frac{M_D^{2+n}}{(2\pi)^n} \ .
\end{equation}
In the most straightforward scenarios~\cite{Antoniadis:1998ig} with
flat extra dimensions of equal length, TeV-scale gravity is excluded
for $n=1$ by solar system tests of Newtonian gravity.  Astrophysical
bounds~\cite{Cullen:1999hc} on supernova cooling and neutron star
heating provide the most stringent bounds for $n=2$ ($\md \agt
600~\tev$) and $n=3$ ($\md \agt 10~\tev$).  For $n\ge 4$, the most
stringent bounds are from collider searches for perturbative graviton
effects~\cite{Giudice:1998ck,Pagliarone:2001ff,Abbott:2000zb} and
cosmic ray bounds on black hole production~\cite{Ringwald:2001vk,%
Anchordoqui:2001cg, Baltrusaitis:mt,agasa}.  These constraints are
each subject to a variety of theoretical assumptions (for a comparison
and discussion, see, \eg, Ref.~\cite{Anchordoqui:2001cg}), but the
most stringent of these require roughly $\md \agt 1~\tev$.

To determine event rates for neutrino telescopes, we must first model
black hole production and evaporation.  For production by cosmic
neutrinos, we follow the analysis of Ref.~\cite{Feng:2001ib}.  Black
holes produced in parton collisions are typically far smaller than the
length scales of the extra dimensions.  These black holes are then
well-approximated by (4+$n$)-dimensional solutions.  The Schwarzschild
radius for a (4+$n$)-dimensional black hole with mass $\mbh$ and
vanishing charge and angular momentum is~\cite{Myers:1986un}
\begin{equation}
r_s (\mbh^2) = \frac{1}{\md}
\left[ \frac{\mbh}{\md} \right]^{\frac{1}{1+n}}
\left[ \frac{2^n \pi^{\frac{n-3}{2}} \Gamma \left(\frac{3+n}{2}
\right)}
{2+n} \right]^{\frac{1}{1+n}} \ .
\end{equation}
We assume that two partons $i$ and $j$ with center-of-mass energy
$\sqrt{\hat{s}}$ form a black hole of mass $\mbh = \sqrt{\hat{s}}$
when they pass within a distance $r_s(\hat{s})$, leading to a
geometric cross section
of~\cite{Banks:1999gd,Giddings:2001bu,Dimopoulos:2001hw}
\begin{equation}
\label{sigmabh}
\hat{\sigma} (ij \to \text{BH}) (\hat{s}) = \pi r_s^2(\hat{s}) \ .
\end{equation}
Evidence from analyses of axisymmetric~\cite{D'Eath:hb} and
off-axis~\cite{Eardley:2002re} classical collisions, an analysis in a
simple model framework~\cite{Solodukhin:2002ui}, and a string
calculation~\cite{Dimopoulos:2001qe} suggests that this picture is
valid semi-classically and is not subject to large
corrections~\cite{Giddings:2001ih}.  Modifications for non-vanishing
angular momentum and spinning black holes have also been found to be
small~\cite{Anchordoqui:2001cg}. However, Voloshin has
argued~\cite{Voloshin:2001vs,Voloshin:2001fe} that the cross section
could be suppressed by the factor $e^{-I}$, where the action is
\begin{equation}
I= \frac{S}{n+1} = \frac{4\pi \, \mbh\, r_s}{(n+1)(n+2)} \ ,
\end{equation}
with $S$ the black hole entropy. This implies vanishing cross sections
in the classical limit, contrary to the evidence noted above.  To
explore the impact of modifications to the black hole production cross
section, however, we consider cases both with and without this
suppression.

The neutrino-nucleon scattering cross section is
then~\cite{Feng:2001ib}
\begin{equation}
\sigma ( \nu N \to \text{BH}) = \sum_i \int_{(\xmin \md)^2/s}^1 dx\,
\hat{\sigma}_i ( xs ) \, f_i (x, Q) \ ,
\label{cross}
\end{equation}
where 
\begin{equation}
\xmin \equiv \mbhmin / \md \agt 1
\end{equation}
parameterizes the minimal black hole mass for which we expect a
semi-classical description to be valid, $s = 2 m_N E_{\nu}$, the sum
is over all partons $i$, and $f_i$ are the parton distribution
functions. The cross section is highly insensitive to choice of
momentum transfer~\cite{Feng:2001ib}; \eg, the choices $Q =
\mbh$~\cite{Dimopoulos:2001hw} and $Q =
r_s^{-1}$~\cite{Emparan:2001kf}, lead to cross section differences of
only 10\% to 20\%.  These cross sections are typically also
insensitive to uncertainties at low $x$.  For example, even extremely
high energy neutrinos with $E_{\nu} \sim 10^{10}~\gev$ probe only $x
\agt (1~\tev)^2 / 10^{10}~\gev^2 \approx 10^{-4}$, within the range of
validity of the CTEQ5 parton distribution functions we
use~\cite{Lai:2000wy}.

\begin{figure}[tbp]
\begin{minipage}[t]{0.49\textwidth}
\postscript{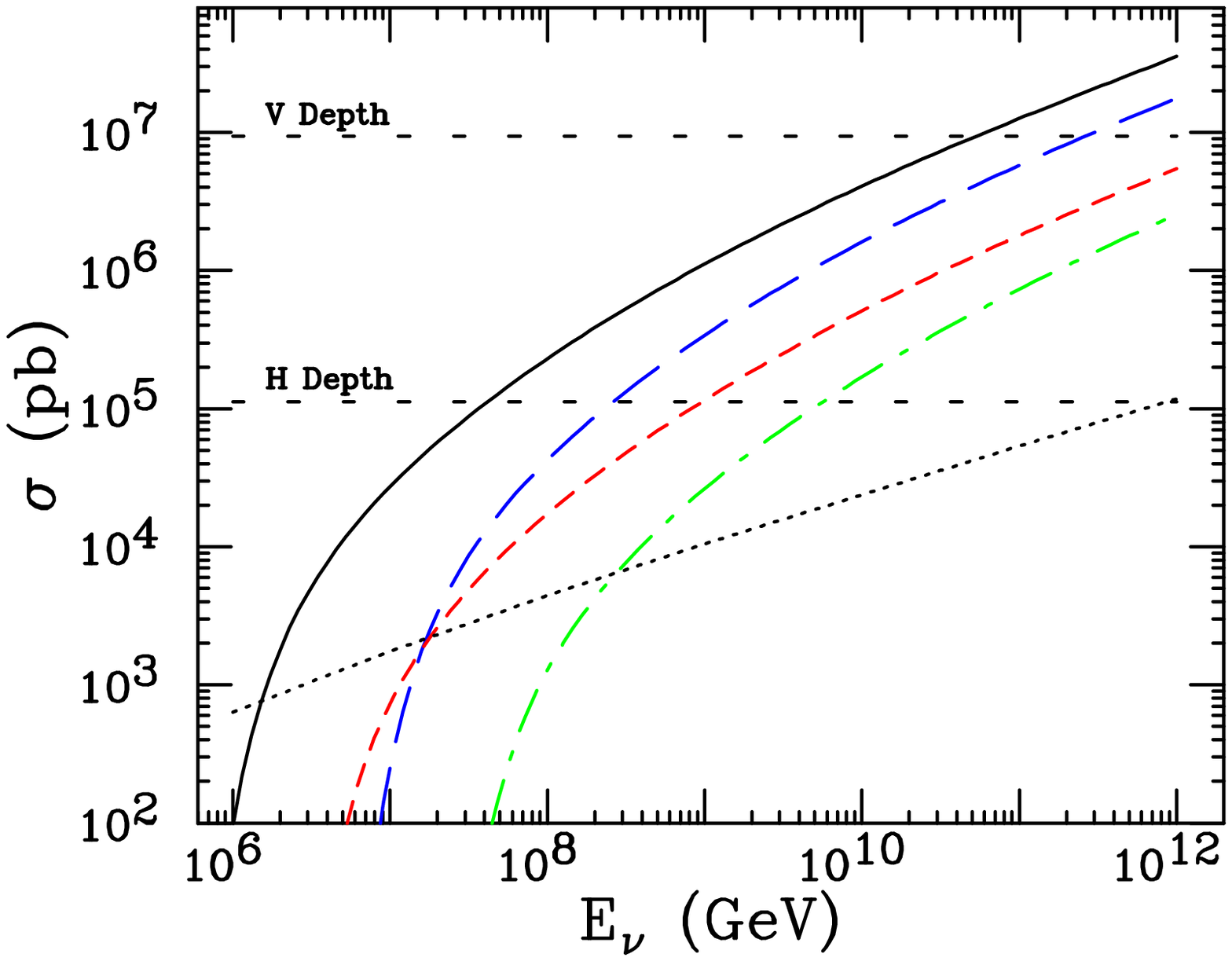}{0.99}
\end{minipage}
\hfill
\begin{minipage}[t]{0.49\textwidth}
\postscript{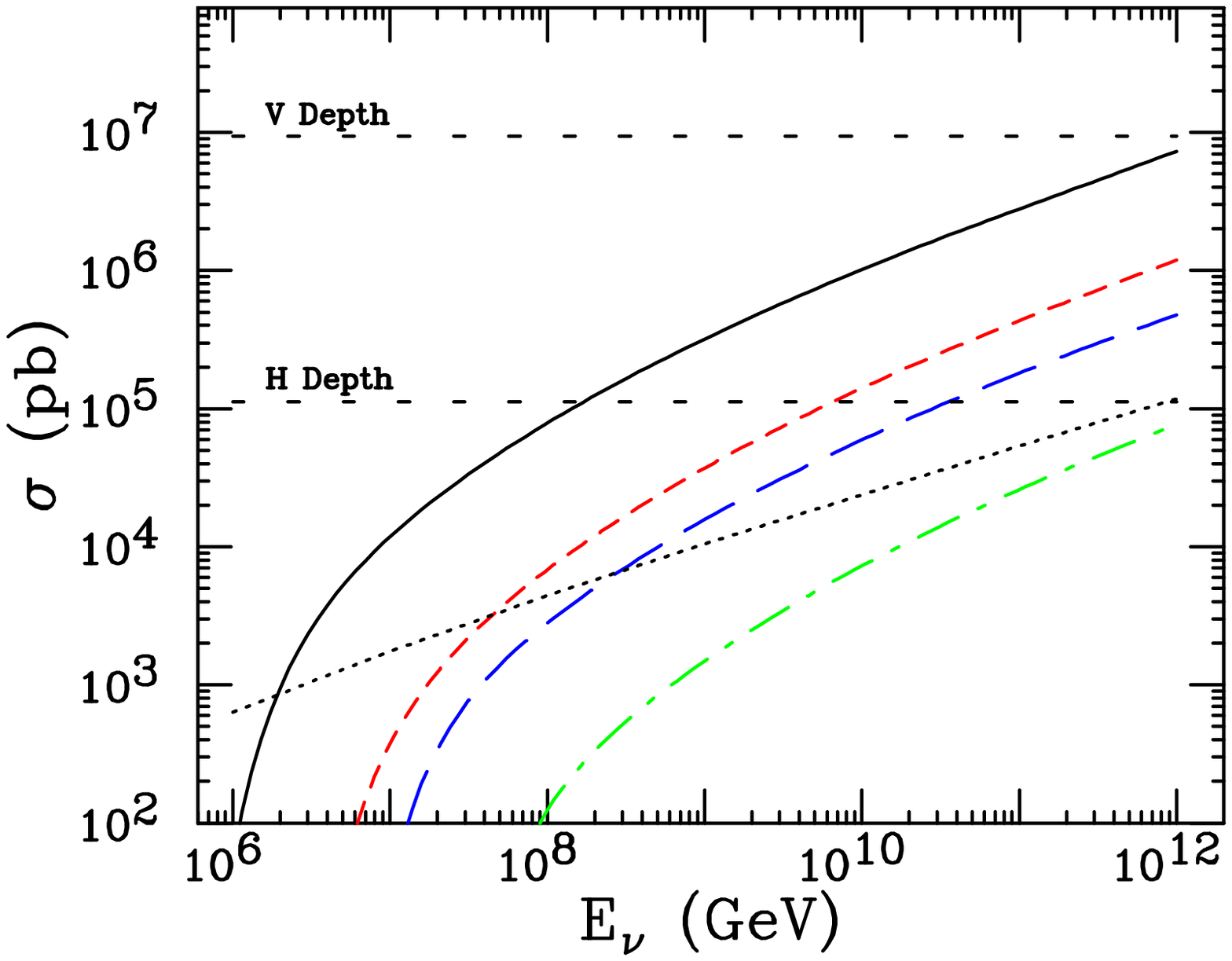}{0.99}
\end{minipage}
\caption{Cross sections $\sigma(\nu N \to {\rm BH})$ for $n=6$ and
$(\md, \xmin) = (1~\tev, 1)$ (solid), $(1~\tev, 3)$ (long dash),
$(2~\tev, 1)$ (short dash), $(2~\tev, 3)$ (dot-dash), and parton cross
sections $\pi r_s^2$ (left) and $\pi r_s^2 e^{-I}$ (right). The dotted
curve is the SM cross section $\sigma(\nu N \to \ell X)$.  The
horizontal lines are the cross sections corresponding to interaction
lengths equal to the vertical and horizontal depths of IceCube. (See
text.)
\label{fig:sigma}
}
\end{figure}

Cross sections for black hole production by cosmic neutrinos are given
in Fig.~\ref{fig:sigma}. The SM cross section for $\nu N \to \ell X$
is included for comparison (dotted curve).  As a result of the sum
over all partons and the lack of suppression from small perturbative
couplings, the black hole cross section may exceed SM interaction
rates by two or more orders of magnitude.

Note that in our conventions, the cross section rises for increasing
$n$ and fixed $\md$.  In conventions where the fundamental Planck
scale is taken to be $\mstar$ with
\begin{equation}
\frac{1}{G_D} = \mstar^{2+n} \ ,
\end{equation}
this behavior is reversed: the cross section decreases for increasing
$n$ and fixed $\mstar$~\cite{Feng:2001ib}.  The dependence of the
cross section on $n$ for fixed Planck scale is convention-dependent
and unphysical; we have adopted the $\md$ convention to simplify
comparison with existing collider bounds.

Assuming a constant density of $1.0~\g/\cm^3$ for the Earth's surface
near the detector, as is valid for IceCube, the neutrino's interaction
length in Earth is
\begin{equation}
L = 1.7\times 10^7~\km \left( \frac{\pb}{\sigma} \right) \ .
\end{equation}
The center of IceCube is at a depth of roughly 1.8 km. A neutrino
reaching this point horizontally passes through 150 km of Earth.  The
cross sections corresponding to neutrino interaction lengths equal to
these two lengths, that is, the horizontal and vertical depths of
IceCube, are also given in Fig.~\ref{fig:sigma}.  We see that for the
geometric cross section, $\md \sim 1~\tev$, and neutrino energies
$E_{\nu} \sim 10^9~\gev$ where the cosmogenic flux peaks (see
Sec.~\ref{sec:fluxes}), black hole production increases the
probability of conversion in down-going neutrinos without increasing
the cross section so much that vertical neutrinos are shadowed by the
Earth.  We therefore expect significantly enhanced rates in neutrino
telescopes.  (On the other hand, the upgoing event rates will be even
more suppressed.) For the exponentially suppressed cross section, a
similar enhancement is also possible for $\md = 1$ to $2~\tev$.

Once produced, these tiny black holes evaporate with rest lifetime of
order $\tev^{-1} \sim 10^{-27}~\s$.  Even though highly boosted, they
decay before accreting matter.  They evaporate in a thermal
distribution with temperature $T_H = (1+n)/(4\pi
r_s)$~\cite{Hawking:1975sw,Myers:1986un} and average
multiplicity~\cite{Giddings:2001bu,Dimopoulos:2001hw}
\begin{equation}
\langle N \rangle \approx \frac{\mbh}{2 T_H} =
\frac{2 \pi}{1+n}
\left[ \frac{\mbh}{\md} \right]^{\frac{2+n}{1+n}}
\left[ \frac{2^n \pi^{\frac{n-3}{2}} \Gamma \left(\frac{3+n}{2}
\right)}
{2+n} \right]^{\frac{1}{1+n}} \ .
\label{multiplicity}
\end{equation}
Neglecting particle masses, the decay products are distributed
according to the number of degrees of freedom~\cite{Emparan:2000rs}:
quarks (72), gluons (16), charged leptons (12), neutrinos (6), $W$ and
$Z$ bosons (9), photons (2), Higgs bosons (1), and gravitons (2).  We
neglect the possibility of other low mass degrees of freedom, such as
right-handed neutrinos and supersymmetric particles. About 75\% of the
black hole's energy is radiated in hadronic degrees of freedom, while
the probability of any given decay particle being a muon (or a tau) is
approximately 3\%.

\begin{figure}[tbp]
\begin{minipage}[t]{0.49\textwidth}
\postscript{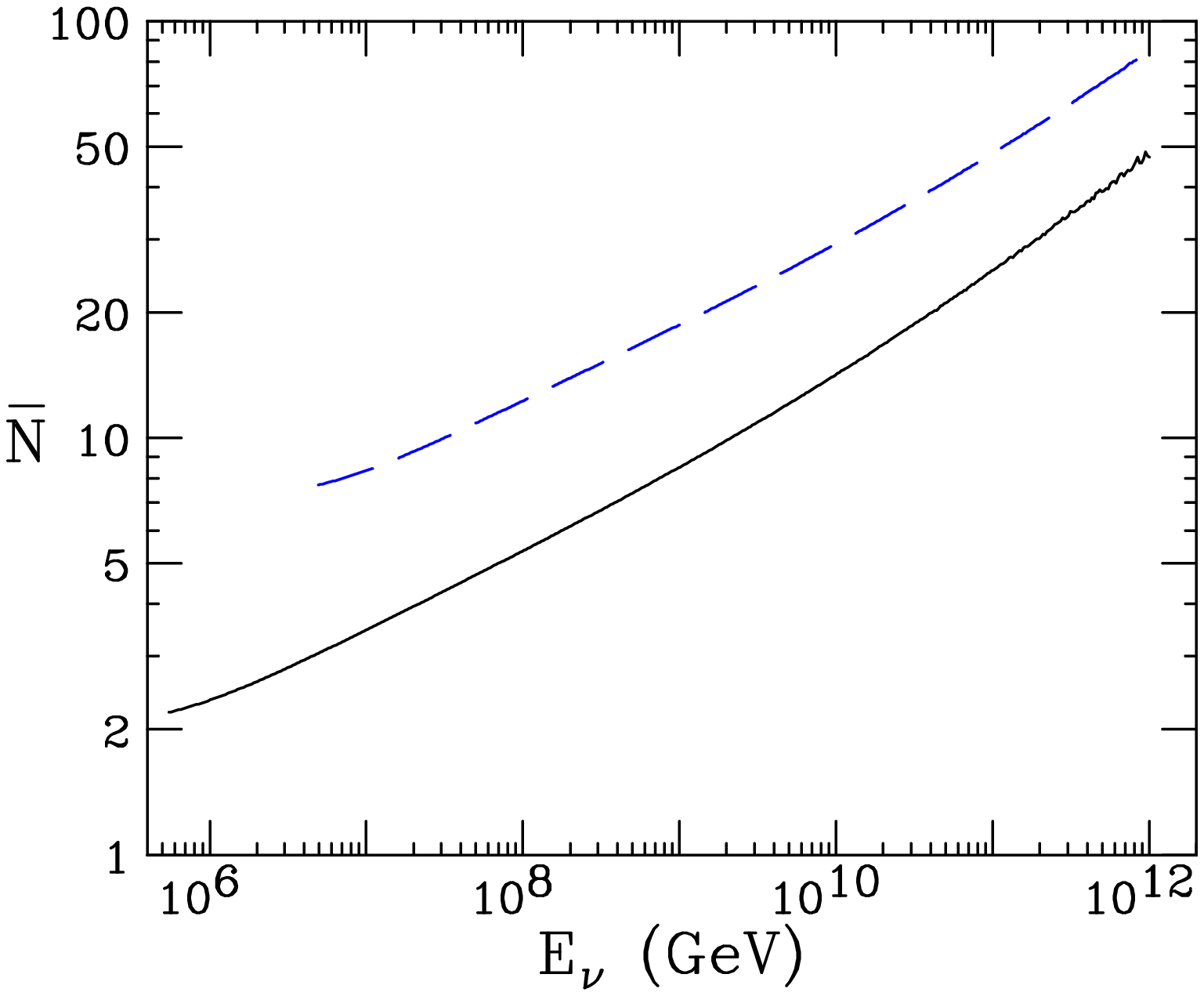}{0.99}
\end{minipage}
\hfill
\begin{minipage}[t]{0.49\textwidth}
\postscript{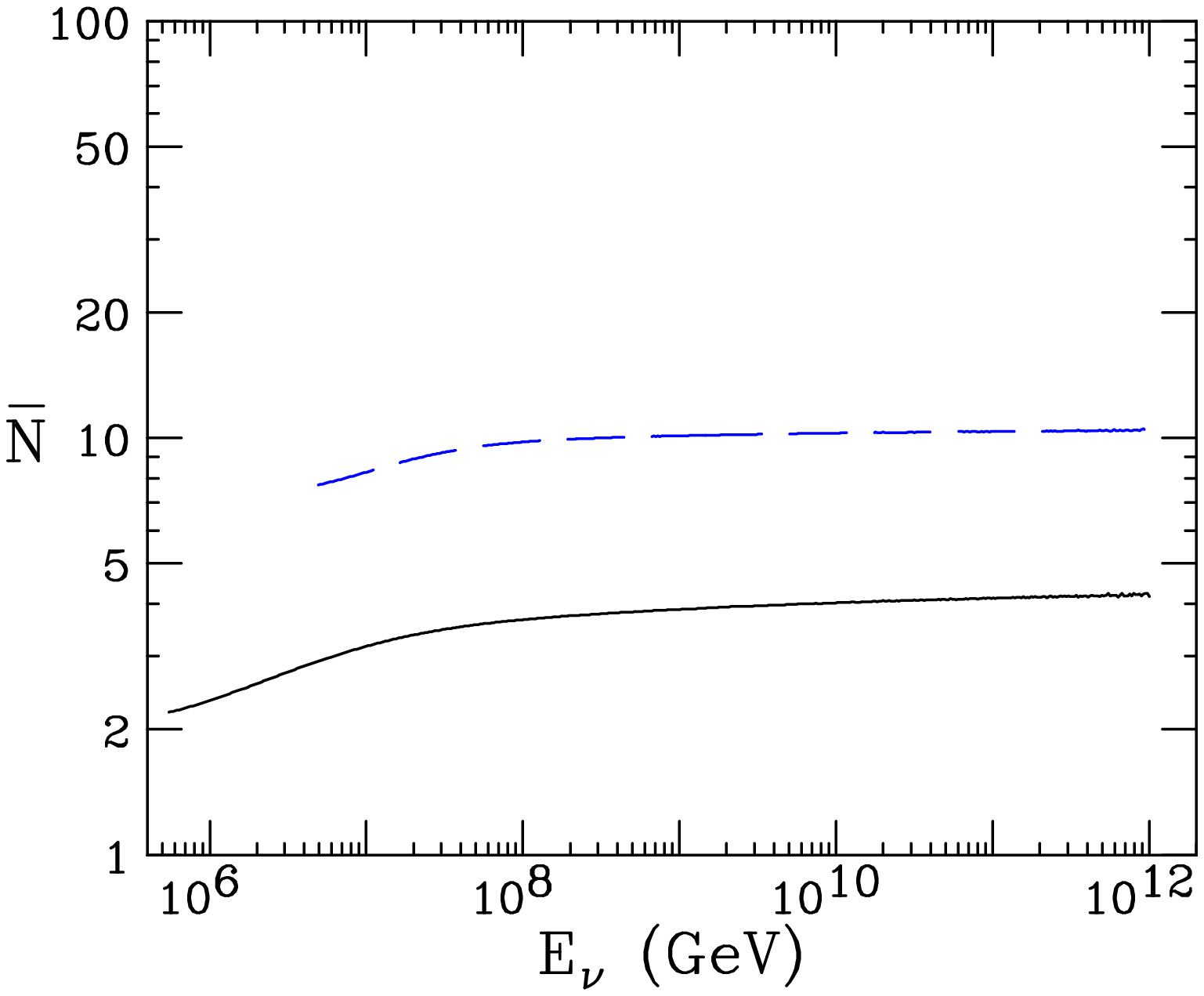}{0.99}
\end{minipage}
\caption{Weighted multiplicities $\overline{N}$ for $n=6$,
$\md=1~\tev$, and $\xmin = 1$ (solid) and 3 (long dash), and parton
cross sections $\pi r_s^2$ (left) and $\pi r_s^2 e^{-I}$ (right).
\label{fig:mult}
}
\end{figure}

As through-going muons and taus will be a promising signal in neutrino
telescopes, the typical multiplicity of black hole decays is of great
importance.  To quantify this, we define a weighted multiplicity
\begin{equation}
\overline{N} = 
\frac{1}{\sigma} \sum_i \int_{(\xmin \md)^2/s}^1 dx\,
\langle N \rangle \, \hat{\sigma}_i ( xs ) \, f_i (x, Q) \ ,
\label{averageN}
\end{equation}
where $\sigma$ is given in \eqref{cross}, and $\langle N \rangle$ is
as in \eqref{multiplicity} with $\mbh^2 = xs$.  The weighted
multiplicity $\overline{N}$ is given in Fig.~\ref{fig:mult} for
various cases with $n=6$ and $\md=1~\tev$.  For the geometric black
hole cross section, these multiplicities may be substantially enhanced
for lower $n$; for the exponentially suppressed cross section, the
dependence on $n$ is slight.  We find that $\overline{N} \sim {\cal
O}(10)$ is possible for ultra-high energy neutrinos.  Note that while
raising $\xmin$ and including the exponential suppression factor both
suppress the total cross section, they have opposite effects on
$\overline{N}$: raising $\xmin$ eliminates events with relatively low
multiplicity, and so raises $\overline{N}$, while the exponential
suppression is largest for events with large $\mbh$ and large
multiplicity, and so suppresses $\overline{N}$.

\section{Astrophysical Sources of Ultra-High Energy Neutrinos}
\label{sec:fluxes}

Observations of black hole formation and decay are most easily made
with neutrino fluxes that are large near or above the EeV scale.
Several such sources have been proposed in the literature.  While a
detailed discussion of these sources is beyond the scope of this
paper, we briefly describe here a few of them, including those used in
our study.

\begin{figure}[tbp]
\postscript{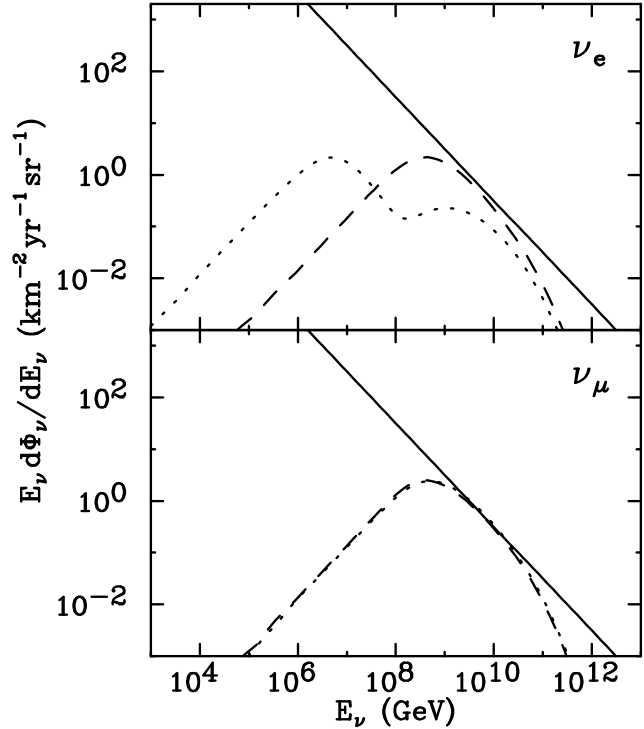}{0.51}
\caption{Representative fluxes: cosmogenic $\nu$ (dashed) and
$\bar{\nu}$ (dotted)~\protect\cite{Engel:2001hd} and the
Waxman-Bahcall flux $\Phi_{\nu_e} = \Phi_{\nu_{\mu}} =
\Phi_{\bar{\nu}_{\mu}}$ (solid)~\protect\cite{Waxman:1998yy}.
\label{fig:engel-nu-CMB}
}
\end{figure}

First, neutrinos are almost certainly produced through pion decays in
the scattering of protons off the cosmic microwave background, $p
\gamma_{\rm CMB} \to n \pi^+$~\cite{Stecker:1979ah}.  This cosmogenic
flux is subject to a number of quantitative uncertainties, including
cosmological source evolution.  As representative fluxes, we consider
the recent results presented in Fig.~4 of Ref.~\cite{Engel:2001hd}.
These fluxes are shown in Fig.~\ref{fig:engel-nu-CMB}.

Second, gamma ray bursts have also been considered as a possible
source of the highest energy cosmic rays.  If this is the case, Fermi
accelerated protons from shocks will generate extremely high energy
neutrinos with energy spectrum $d\Phi_\nu/dE_{\nu} \propto
E_{\nu}^{-2}$~\cite{Waxman:1997ti, Waxman:2001tk, Halzen:1999xc,
Alvarez-Muniz:2000st}.  This neutrino flux, as well as those from
other compact sources, such as active galactic nuclei, is limited by
the Waxman-Bahcall (WB) bound~\cite{Waxman:1998yy}.  This constraint
is valid for all astrophysical neutrino sources that are optically
thin to $p\gamma$ and $pp$ interactions.  We consider a conservative
estimate of this bound, $E_\nu^2\, d\Phi_\nu/dE_{\nu} = 1\times
10^{-8}~\cm^{-2}~\s^{-1}~\sr^{-1}$~\cite{Waxman:1998yy}, where here
$\nu = \nu_e, \nu_{\mu}, \bar{\nu}_{\mu}$. This flux is also shown in
Fig.~\ref{fig:engel-nu-CMB}.  It is approximately equal to the
cosmogenic flux for $E_{\nu} \sim 10^{10}~\gev$, but is much larger
for lower (and higher) energies.

Third, if the highest energy cosmic rays observed are generated by the
annihilation of superheavy dark matter particles or from the decay of
topological defects, neutrinos will also be produced.  Such fluxes
have been described in Refs.~\cite{Halzen:2001ec,Sarkar:2001se,%
Birkel:1998nx,Berezinsky:1997hy}.  We will not discuss these sources
further, but note that they may predict large neutrino fluxes at
extremely high energies, enhancing the results given below.

In propagating to the Earth, the neutrino fluxes of
Fig.~\ref{fig:engel-nu-CMB} will mix~\cite{Athar:2000yw}.  Given the
solutions preferred by neutrino oscillation experiments and the
enormous distances traveled, we take the neutrinos that reach the
Earth to be in the ratio $\nu_e : \nu_{\mu} : \nu_{\tau} = 1:1:1$ and
similarly for anti-neutrinos.

\section{Event Simulation}

With regard to signals at neutrino telescopes, black hole evaporation
products may be divided into three categories: showers (hadronic and
electromagnetic), muons, and taus.  Black holes also decay to
neutrinos, but we neglect this flux in this work.

The ranges of typical hadronic and electromagnetic showers are much
less than the linear dimension of large-scale neutrino telescopes, and
so, to first approximation, only contained showers from black holes
produced inside the detector may be detected.  The backgrounds for
showers from black hole evaporation consist of hadronic showers from
neutral and charged current neutrino events and electromagnetic
showers from charged current electron neutrino events.  At IceCube,
hadronic shower energies should be measurable with an accuracy of
about 30\%.

In contrast to showers, at the typical energies of black hole events,
muons travel several kilometers before losing a decade of energy.  The
dominant signal is therefore through-going muons.  IceCube can measure
the energy and direction of any observed muon. The angular resolution
is about $2^\circ-3^\circ$ while the energy resolution is
approximately a factor of three. Signal and background muons may
therefore be differentiated with an energy cutoff.  As we will see,
for down-going muons with energy above 500 TeV to 1 PeV, the black
hole signal may be well above the SM background from atmospheric
neutrinos. Note that for black holes produced sufficiently near the
detector, muon events may be obscured by showers.  This occurrence is
rare, however, and we ignore this possibility below.

At high energies, when tau decay is sufficiently time dilated, taus
have ranges as large or larger than muons, and so the dominant tau
signal is from through-going taus.  These events have a characteristic
signature consisting of a ``clean'' track, \ie, a track without much
energy lost through low energy cascades.  We will consider all events
in which a tau track passes through the detector.  Taus can be
differentiated from muons by the cleanliness of their tracks or
through their decays in the detector leading to ``lollipop''
events~\cite{IceCubetaus}, described later in this paper.  We assume
that taus and muons are distinguishable at all energies, although it
will be difficult to distinguish a slow muon of energy $\alt 200~\gev$
from a very high energy tau, as they can both generate clean,
through-going tracks in the detector.

Very massive black holes with large multiplicity decays may evaporate
to several muons (or taus).  However, these travel in coincidence
through the detector.  Their angular spread is only $\Delta \theta
\sim T_H \langle N \rangle / E_{\nu}$, much less than a degree, and so
cannot be resolved with large Cerenkov detectors such as IceCube.
Spectacular multi-lepton signatures are therefore not possible.  Note,
however, that more massive black holes are easier to detect, as they
produce leptons with greater energy, which travel further before
dropping below the cutoff energy.

To evaluate black hole detection prospects, it is essential to
determine, in a unified framework, both the SM background and the
black hole event rate.  We now describe both of these calculations.

\subsection{Standard Model Events}
\label{sec:SM}

\subsubsection{Showers}

In the SM, a general expression for the total number of shower events
in an underground detector is
\begin{eqnarray}
N_{\rm sh} &=& \sum _{i,j} 2\pi A T \int d\cos\theta_z 
\int dE_{\nu_i} \frac{d\Phi_{\nu_i}}{dE_{\nu_i}}(E_{\nu_i}) \,
P_{\rm surv} \nonumber \\
&& \times\int_{y^{i,j}_{\rm min}}^{y^{i,j}_{\rm max}} dy 
\frac{1}{\sigma_{\rm SM}^j(E_{\nu_i})} \,
\frac{d\sigma_{\rm SM}^j}{dy}(E_{\nu_i}) \, P_{\rm int} \ ,
\label{rate}
\end{eqnarray}
where $\theta_z$ is the zenith angle ($\theta_z = 0$ is vertically
downward), and the sums are over neutrino (and anti-neutrino) flavors
$i=e, \mu, \tau$ and interactions $j= {\rm CC}$ (charged current) and
NC (neutral current).  $A$ is the detector's cross sectional area with
respect to the $\nu$ flux, $T$ is its observation time, and
$d\Phi_{\nu_i}/dE_{\nu_i}$ is the differential neutrino flux that
reaches the Earth.  For $i=\tau$, \eqref{rate} is modified to include
the effects of regeneration, as discussed below.

$P_{\rm surv}$ is the probability that a neutrino survives to reach
the detector.  It is given by
\begin{equation} 
P_{\rm surv} \equiv 
\exp [-X(\theta_z)\sigma^{\rm tot}_{\rm SM}(E_{\nu_i}) N_A] \ ,
\label{survival}
\end{equation}
where $N_A \simeq 6.022 \times 10^{23}~\g^{-1}$, and the total
neutrino interaction cross section is
\begin{equation}
\sigma^{\rm tot}_{\rm SM} = \sigma^{\rm CC} + \sigma^{\rm NC} \ .
\end{equation}
Note that this is conservative, as it neglects the possibility of a
neutrino interacting through a NC interaction and continuing on to
create a contained shower.  $X(\theta_z)$ is the column density of
material the neutrino must traverse to reach the detector with zenith
angle $\theta_z$. It depends on the depth of the detector and is given
by
\begin{equation}
X(\theta_z)=\int_{\theta_z} \rho(r(\theta_z,l)) \, dl \ ,
\end{equation} 
the path length along direction $\theta_z$ weighted by the Earth's
density $\rho$ at distance $r$ from the Earth's center.  For the
Earth's density profile, we adopt the piecewise continuous density
function $\rho(r)$ of the Preliminary Earth Model~\cite{preliminary}.

$P_{\rm int}$ is the probability that the neutrino interacts in the
detector. It is given by
\begin{equation}
P_{\rm int} = 1 - \exp \left[ - \frac{L}{L_{\rm SM}^j (E_{\nu_i})}
\right] \ ,
\end{equation}
where, for showers, $L$ is the linear dimension of the detector, and
$L_{\rm SM}^j (E_{\nu_i})$ is the mean free path for neutrino
interaction of type $j$.  For realistic detectors, $L \ll L_{\rm SM}^j
(E_{\nu_i})$, and so $P_{\rm int} \approx L/L_{\rm SM}^j (E_{\nu_i})$.
To an excellent approximation, then, the shower event rate scales
linearly with detector volume $V=AL$, and we present results in units
of events/volume/time.

Finally, the inelasticity parameter $y$ is the fraction of the
neutrino energy carried away in hadrons.  The limits of integration
are determined by the interaction type and neutrino flavor.  For NC
$\nu_e$ interactions and all $\nu_\mu$ and $\nu_\tau$ interactions,
$y_{\rm max}=1$ and $y_{\rm min}=E^{\rm thr}_{\rm sh}/E_\nu$, where
$E^{\rm thr}_{\rm sh}$ is the threshold energy for shower
detection. For CC $\nu_e$ interactions, the outgoing electron also
showers, and so $y_{\rm max}=1$ and $y_{\rm min}=0$.

\subsubsection{Muons}
\label{sec:muons}

Energetic through-going muons are produced only by $\nu_\mu$ CC
interactions.  For a muon to be detected, it must reach the detector
with energy above some threshold $E^{\rm thr}_\mu$.  The expression of
\eqref{rate} then also describes the number of muon events with
$P_{\rm surv}$ as before, but with
\begin{equation}
P_{\rm int} = 1-\exp\left[-\frac{R_\mu}
{L^{\rm CC}_{\rm SM}(E_{\nu_\mu})} \right] \ ,
\end{equation}
where $R_\mu$ is the range of a muon with initial energy $E_\mu =
(1-y) E_{\nu_\mu}$ and final energy $E^{\rm thr}_\mu$. We assume muons
lose energy continuously according to
\begin{equation}
\frac{dE}{dX}=-\alpha - \beta E \ ,
\end{equation}
where $\alpha=2.0~\mev~\cm^2/\g$ and $\beta=4.2 \times
10^{-6}~\cm^2/\g$~\cite{Dutta:2000hh}. The muon range is then
\begin{equation}  
R_\mu = \frac{1}{\beta} \ln \left[ 
\frac{\alpha + \beta E_\mu}{\alpha + \beta E^{\rm thr}_\mu} \right]
\ .
\label{murange}
\end{equation}
In this case, $y_{\rm max} = 1 - E^{\rm thr}_\mu/E_\nu$ and $y_{\rm
min}=0$.

The event rate for muons is significantly enhanced by the possibility
of muons propagating from several km into the detector.  Note,
however, that this enhancement is $\theta_z$-dependent: for nearly
vertical down-going paths, the path length of the muon is limited by
the amount of matter above the detector, not by the muon's range.
This is taken into account explicitly in the simulations, and its
effect will be evident in the results presented in
Sec.~\ref{sec:distributions}.

At extremely high neutrino energies, the approximate form $P_{\rm int}
\approx R_\mu/L^{\rm CC}_{\rm SM}(E_{\nu_\mu})$, often presented in
the literature, is less accurate than the one used here, and the
difference may be significant in scenarios in which the neutrino cross
section is enhanced with respect to the SM value, such as the ones we
explore in this paper.  Note also that in this case, our expression
for $P_{\rm surv}$ is conservative, as it demands that the neutrino
survive all the way to the detector, neglecting the possibility that
it may convert a significant distance from the detector and still
produce a signal.

\subsubsection{Taus}

Taus are produced only by CC $\nu_\tau$ interactions. This process
differs significantly from the muon case, as tau neutrinos are
regenerated by tau decay through $\nu_\tau \to \tau \to
\nu_\tau$~\cite{Halzen:1998be}. As a result, for tau neutrinos, CC and
NC interactions do not deplete the $\nu_\tau$ flux, but serve only to
soften the neutrino energy. We include this important effect by first
performing a dedicated simulation that determines
$\overline{E}_{\nu_\tau} (E_{\nu_\tau}, \theta_z)$, the average energy
a $\nu_\tau$ has when it reaches the detector, as a function of its
initial energy $E_{\nu_\tau}$ and zenith angle $\theta_z$.  The tau
event rate is then given by
\begin{eqnarray}
N_{\tau} &=& 2\pi A T \int d\cos\theta_z
\int dE_{\nu_\tau} \frac{d\Phi_{\nu_\tau}}{dE_{\nu_\tau}}(E_{\nu_\tau})
\int_{y_{\rm min}}^{y_{\rm max}} dy
\frac{1}{\sigma_{\rm SM}^{\rm CC}(\overline{E}_{\nu_\tau})}
\frac{d\sigma_{\rm SM}^{\rm CC}}{dy}(\overline{E}_{\nu_\tau})
\nonumber \\
&& \times \left[1- \exp \left(-\frac{R_\tau((1-y)
\overline{E}_{\nu_\tau})}
{L^{\rm CC}_{\rm SM}(\overline{E}_{\nu_\tau})}\right)\right]
\Theta ((1-y)\overline{E}_{\nu_{\tau}} - E^{\rm thr}_\tau) \ , 
\label{taurate}
\end{eqnarray}
where $R_\tau((1-y) \overline{E}_{\nu_\tau})$ is the range of the
produced tau, but evaluated at the energy of the tau neutrino after
regeneration.  $R_\tau$ is given by \eqref{murange} but now with
$\beta=3.6 \times 10^{-7}~\cm^{-2}/\g$~\cite{Dutta:2000hh}.  The last
factor takes into account the requirement that the tau track be long
enough to be identified in the detector.  We require $E^{\rm thr}_\tau
\simeq 2.5 \times 10^6~\gev$ so that the tau decay length is above 125
m, the string separation length in IceCube.  It is not clear, at this
time, whether through-going tau events will be separable from less
energetic muon events.  Those tau events that include one (lollipop
events) or two (double bang events) showers in the detector volume
will be identifiable, however.  The rate of down-going lollipop events
in a $\km^3$ neutrino telescope is expected to be of the order of the
rate of down-going shower events, probably slightly smaller.  Double
bang events will be mostly observed for neutrino energies in a limited
range between roughly 10 and 100 PeV~\cite{IceCubetaus}.  Our results
assume that all tau events can be distinguished from muon events,
although this may be difficult to realize.

As with muons, at very high energies taus can travel several
kilometers before decaying or suffering significant energy loss.  The
enhancement to tau event rates from this effect is
$\theta_z$-dependent as discussed above for muons.

\subsection{Black Hole Events}

Black hole event rates may be determined with only minor
modifications.  For showers, the corresponding expression is
\begin{eqnarray}
N_{\rm BH} &=& \sum_i 2\pi A T \int d\cos\theta_z 
\int dE_{\nu_i} \frac{d\Phi_{\nu_i}}{dE_{\nu_i}}(E_{\nu_i}) \,
P_{\rm surv} \nonumber \\
&& \times\int_0^1 dy 
\frac{1}{\sigma_{\rm BH}(E_{\nu_i})} \,
\frac{d\sigma_{\rm BH}}{dy}(E_{\nu_i}) \, P_{\rm int} 
P_{\rm BH}(E_{\nu_i}) \ .
\label{BHrate}
\end{eqnarray}
The survival and interaction probabilities are now
\begin{eqnarray} 
P_{\rm surv} &=& \exp [-X(\theta_z)\sigma^{\rm tot}(E_{\nu_i}) 
N_A] \\
P_{\rm int} &=& 1-\exp\left[-\frac{L}
{L_{\rm BH}(E_{\nu_i})} \right] \ ,
\label{survival2}
\end{eqnarray}
where
\begin{equation}
\sigma^{\rm tot}=\sigma^{\rm tot}_{\rm SM} + \sigma_{\rm BH} \ ,
\end{equation}
with $\sigma_{\rm BH}$ the cross section for black hole production,
and $L_{\rm BH}(E_{\nu_i})$ is the neutrino mean free path for black
hole production.  As in the SM, $L$ is the linear dimension of the
detector in the case of shower events. We assume $d\sigma_{\rm BH}/dy
\propto \delta(y-0.75)$, \ie, that 75\% of the black hole energy is
carried away by showers, and impose that the generated shower has
energy above $E^{\rm thr}_{\rm sh}$ to be detected.  $P_{\rm
BH}(E_\nu) = 1$ for shower events.

In the case of muon and tau events, $L$ is the corresponding range for
the average muon or tau energy. We assume $E_{\mu,\tau}
=E_\nu/\overline{N}$, where $\overline{N}$ is the weighted
multiplicity discussed in \eqref{averageN}.  The true $E_{\mu,\tau}$
distribution is essentially flat with endpoints 0 and
$2E_\nu/\overline{N}$.  Our simplification is valid except for lepton
energies near threshold.  In addition, spreading the lepton energy has
two compensating effects, as some leptons below threshold become
detectable, and some above threshold drop below threshold.  We have
checked that the error made is insignificant at the $\sim 10\%$ level.

To account for the branching fraction to leptons, we include
\begin{equation}
P_{\rm BH}(E_{\nu_i}) = 1-{\rm exp}\left[
- \frac{\overline{N}(E_{\nu_i})}{30} \right] \ ,
\end{equation} 
the probability of obtaining at least one muon (or tau) in the decay
of a black hole when the expected muon (or tau) multiplicity is
$\overline{N}/30$.

\section{Rates}

The integrated event rates for showers, muons, and taus are shown in
Tables~\ref{table:I}, \ref{table:II}, and \ref{table:III},
respectively.  We consider IceCube, with a representative depth of
$1.8~\km$. We give results for various $n$, $\md$, $\xmin$, and with
and without exponential suppression in the parton cross section.  Only
down-going rates are presented.  Up-going rates are, as expected,
extremely suppressed in the presence of black hole production, as will
be seen in Sec.~\ref{sec:distributions}.

\begin{table}[tbp]
\caption{Event rate for down-going showers (in $2\pi$ sr) with $E^{\rm
thr}_{\rm sh} = 500~\tev$ in IceCube.  We consider the
Waxman-Bahcall~\protect\cite{Waxman:1998yy} and
cosmogenic~\protect\cite{Engel:2001hd} fluxes, $\md =1$ and 2 TeV, and
various cases $(n,\xmin, \hat{\sigma})$, where $n$ is the number of
extra dimensions, $\xmin \equiv \mbhmin/\md$, and $\hat{\sigma}$ is
the parton level cross section for black hole production.
\label{table:I}
}
\begin{tabular}{ c|c c c c c c } \hline
& \multicolumn{2}{c }{~WB Flux~} & \multicolumn{2}{c }{~Cosmogenic
Flux~} \\ \cline{2-5}
\raisebox{1.7ex}[0pt]{~Showers ($\km^{-3}~\yr^{-1}$)~} 
&~$M_D=1~\tev$~~&~$M_D=2~\tev$~~&~$M_D=1~\tev$~~&~$M_D=2~\tev$~~\\ 
\hline \hline
~~Standard Model~~&~4.8~&~4.8~&~0.1~&~0.1~\\
\hline
~~BH $(6,1,\pi r_s^2)$~~&~44.6~&~3.1~&~5.2~&~0.9~\\
~~BH $(6,3,\pi r_s^2)$~~&~6.5~&~0.5~&~2.3~&~0.3~\\ 
~~BH $(6,1,\pi r_s^2 e^{-I})$~~&~18.5~&~1.2~&~2.1~&~0.3~\\
~~BH $(6,3,\pi r_s^2 e^{-I})$~~&~0.4~&~$2.8\times 10^{-2}$~&
   ~0.1~&~$1.5\times 10^{-2}$~\\ 
\hline
~~BH $(3,1,\pi r_s^2)$~~&~16.3~&~1.1~&~2.7~&~0.4~\\ 
~~BH $(3,3,\pi r_s^2)$~~&~3.0~&~0.2~&~1.2~&~0.1~\\ 
~~BH $(3,1,\pi r_s^2 e^{-I})$~~&~3.8~&~0.2~&~0.5~&
   ~$6.1\times 10^{-2}$~\\ 
~~BH $(3,3,\pi r_s^2 e^{-I})$~~&~$2.9\times 10^{-2}$~&
   ~$1.9\times 10^{-3}$~&~$8.6\times 10^{-3}$~&~$9.3\times 10^{-4}$~\\ 
\hline 
\end{tabular}
\end{table}

\begin{table}[htbp]
\caption{As in Table~\ref{table:I}, but for the flux of down-going
muons and $E^{\rm thr}_{\mu} = 500~\tev$.
\label{table:II}
}
\begin{tabular}{ c|c c c c c c } \hline
& \multicolumn{2}{c }{~WB Flux~} & \multicolumn{2}{c }{~Cosmogenic
Flux~} \\ \cline{2-5}
\raisebox{1.7ex}[0pt]{~Muons ($\km^{-2}~\yr^{-1}$)~} 
&~$M_D=1~\tev$~~&~$M_D=2~\tev$~~&~$M_D=1~\tev$~~&~$M_D=2~\tev$~~\\ 
\hline \hline
~~Standard Model~~&~6.0~&~6.0~&~0.2~&~0.2~\\
\hline
~~BH $(6,1,\pi r_s^2)$~~&~27.7~&~2.6~&~4.9~&~1.1~\\
~~BH $(6,3,\pi r_s^2)$~~&~12.0~&~1.1~&~4.9~&~0.7~\\
~~BH $(6,1,\pi r_s^2 e^{-I})$~~&~8.8~&~0.7~&~1.3~&~0.2~\\
~~BH $(6,3,\pi r_s^2 e^{-I})$~~&~0.6~&~$4.6 \times 10^{-2}$~&~0.2~&
   ~$2.5 \times 10^{-2}$~\\ 
\hline
~~BH $(3,1,\pi r_s^2)$~~&~14.0~&~1.2~&~4.2~&~0.6~\\
~~BH $(3,3,\pi r_s^2)$~~&~7.1~&~0.6~&~3.5~&~0.4~\\
~~BH $(3,1,\pi r_s^2 e^{-I})$~~&~1.8~&~0.1~&~0.3~&
   ~$4.1 \times 10^{-2}$~\\
~~BH $(3,3,\pi r_s^2 e^{-I})$~~&~$4.2 \times 10^{-2}$~&~$3.1 \times
10^{-3}$~&~$1.5 \times 10^{-2}$~&~$1.6 \times 10^{-3}$~\\ 
\hline 
\end{tabular}
\end{table}

\begin{table}[htbp]
\caption{As in Table~\ref{table:I}, but for the flux of down-going
taus and $E^{\rm thr}_{\tau} = 2.5\times 10^6~\gev$.
\label{table:III}
}
\begin{tabular}{ c|c c c c c c } \hline
& \multicolumn{2}{c }{~WB Flux~} & \multicolumn{2}{c }{~Cosmogenic
Flux~} \\ \cline{2-5}
\raisebox{1.7ex}[0pt]{~Taus ($\km^{-2}~\yr^{-1}$)~} 
&~$M_D=1~\tev$~~&~$M_D=2~\tev$~~&~$M_D=1~\tev$~~&~$M_D=2~\tev$~~\\ 
\hline \hline
~~Standard Model~~&~0.9~&~0.9~&~0.1~&~0.1~\\
\hline
~~BH $(6,1,\pi r_s^2)$~~&~15.1~&~2.2~&~5.0~&~1.2~\\
~~BH $(6,3,\pi r_s^2)$~~&~7.7~&~1.0~&~4.5~&~0.7~\\
~~BH $(6,1,\pi r_s^2 e^{-I})$~~&~4.5~&~0.6~&~1.3~&~0.2~\\
~~BH $(6,3,\pi r_s^2 e^{-I})$~~&~0.5~&~$4.4 \times 10^{-2}$~&
   ~0.2~&~$2.9 \times 10^{-2}$~\\ 
\hline
~~BH $(3,1,\pi r_s^2)$~~&~8.7~&~1.1~&~4.1~&~0.7~\\
~~BH $(3,3,\pi r_s^2)$~~&~4.8~&~0.6~&~3.1~&~0.4~\\
~~BH $(3,1,\pi r_s^2 e^{-I})$~~&~1.0~&~0.1~&
   ~0.4~&~$5.2 \times 10^{-2}$~\\
~~BH $(3,3,\pi r_s^2 e^{-I})$~~&~$2.9 \times 10^{-2}$~&~$2.8 \times
   10^{-3}$~&~$1.6 \times 10^{-2}$~&~$1.8 \times 10^{-3}$~\\ 
\hline 
\end{tabular}
\end{table}

For showers, the geometric cross section $\hat{\sigma} = \pi r_s^2$,
and $\md = 1~\tev$, we find a few events per year in each channel for
cosmogenic fluxes, and as many as tens of events per year for the WB
flux.  These event rates are far above SM background.  For the WB
flux, bounds from AGASA and Fly's Eye will imply limits above $\md
=1~\tev$, but even for $\md=2~\tev$, we find reasonable rates.  In
this case, however, the SM background will be important.  For the
exponentially suppressed parton cross section, event rates are
suppressed, but not drastically so.  In fact, for $\xmin=1$, $\sim
{\cal O}(1)$ event per year is possible given the cosmogenic flux.
For $\xmin=3$, the exponential suppression is large, and event rates
are highly suppressed.

Generally, event rates are on more solid footing for larger $\xmin$,
where both the semi-classical production cross section and the
assumption of thermal decay are more reliable.  For showers, however,
our event rate calculation relies essentially only on the requirement
that black holes decay to visible showers and is insensitive to the
details of evaporation.  At the same time, while the production cross
section for black holes (or their stringy Planck mass
progenitors~\cite{Dimopoulos:2001qe}) is subject to significant
quantum corrections for $\mbh \approx \md$, there is no reason to
expect it to vanish or be greatly suppressed.  For showers, then, we
find the requirement $\xmin=1$ reasonable.

Qualitatively similar conclusions apply for muons and taus.  While
these event rates are suppressed relative to shower rates by the
branching ratio for black hole decay to leptons, they are enhanced by
the possibility of muons and taus propagating from many km away into
the detector.  We find that these effects effectively balance each
other.  In the case of muons and taus, we have assumed thermal decay
distributions, which may not be accurate for $\xmin \approx 1$.  The
event rates are hardly reduced for $\xmin =3$, however; as noted in
Sec.~\ref{sec:production}, the more stringent $\xmin$ requirement
preferentially eliminates events with low multiplicities, and so has
little impact on lepton rates.

The muon flux is significantly larger for the WB flux than for the
cosmogenic flux.  This is straightforward to understand --- the WB
flux is larger at low energies.  The tau event rate is not as greatly
enhanced, however.  The tau decay length is $4.9~\km\, (E_{\tau} /
10^8~\gev)$.  While the WB flux enhances fluxes at low energies, the
resulting taus decay before they lose energy, and the tau range is
therefore diminished.  Larger low energy fluxes therefore do not
enhance the tau rate as significantly.

In addition to enhancing the SM event rates by more than an order of
magnitude, Tables~\ref{table:I}-\ref{table:III} show that black hole
production may also change the relative event rates of the various
channels.  The ability of neutrino telescopes, unique among cosmic ray
experiments, to differentiate showers, muons, and possibly taus,
allows one to measure these relative event rates.  Given sufficient
statistics, this may provide an important signal of physics beyond the
SM.

\section{Angle and Energy Distributions}
\label{sec:distributions}

Neutrino telescopes may also measure angle and energy distributions,
providing additional opportunities to distinguish black hole events
from SM or other possible physics and for constraining black hole
properties.

\begin{figure}[tbp]
\postscript{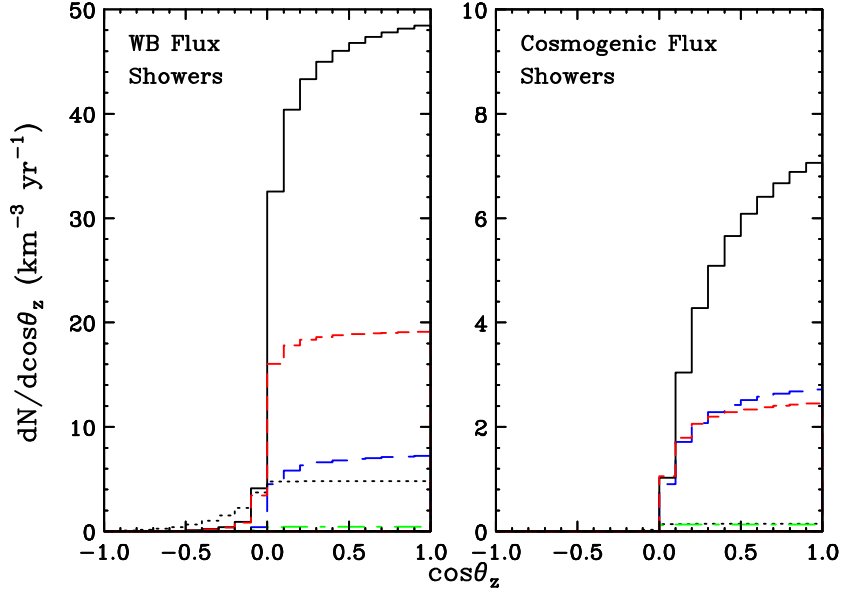}{0.67}
\caption{Zenith angle distribution of shower events for $\md =
1~\tev$, $E^{\rm thr}_{\rm sh} = 500~\tev$, and $(n, \xmin,
\hat{\sigma}) = (6,1,\pi r_s^2)$ (solid), $(6,3,\pi r_s^2)$ (long
dash), $(6,1,\pi r_s^2 e^{-I})$ (short dash), $(6,3,\pi r_s^2 e^{-I})$
(dot-dash). Also shown is the standard model prediction (dotted).
\label{fig:showers_zenith}
}
\end{figure}

\begin{figure}[tbp]
\postscript{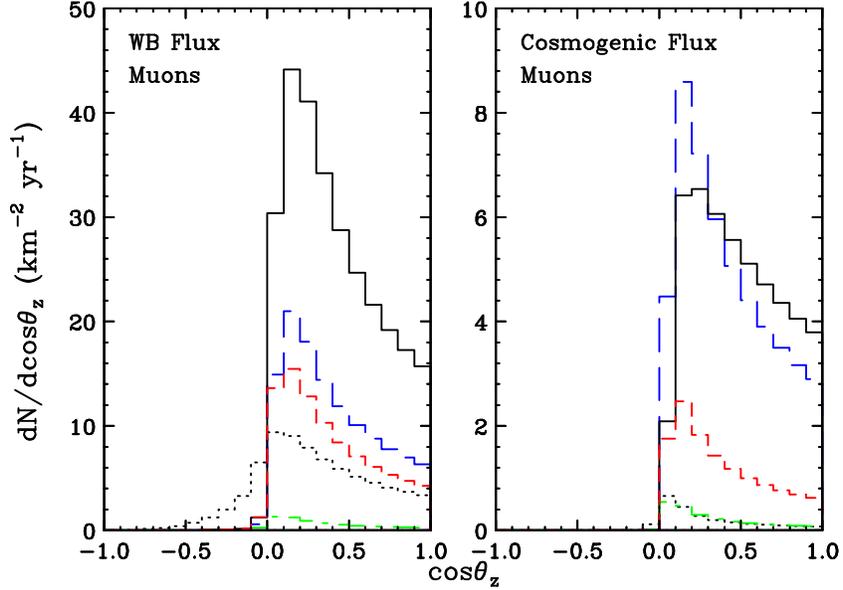}{0.67}
\bigskip
\caption{As in Fig.~\protect\ref{fig:showers_zenith}, but for muon
events with $E^{\rm thr}_{\mu} = 500~\tev$.
\label{fig:muons_zenith}
}
\end{figure}

\begin{figure}[tbp]
\postscript{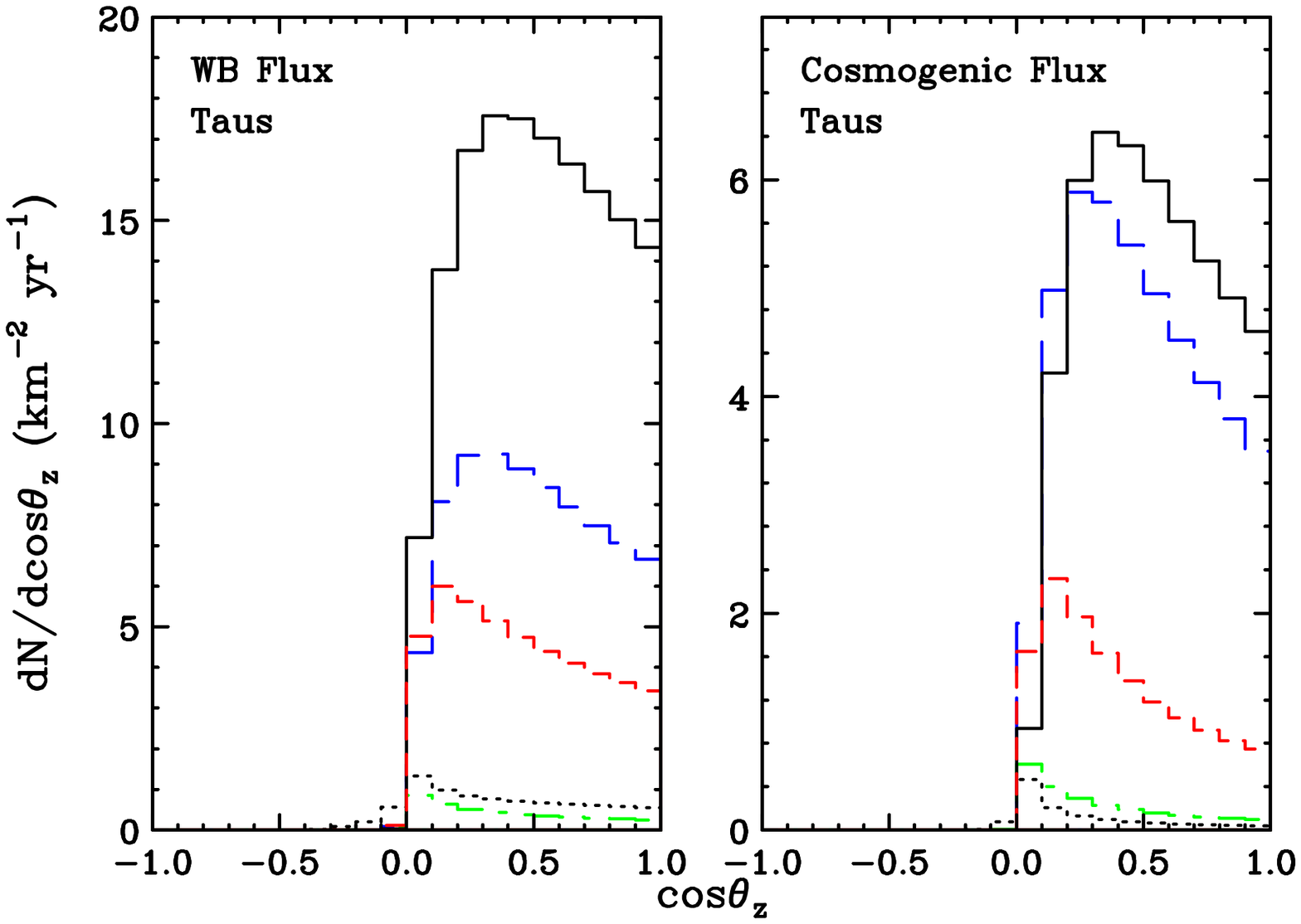}{0.67}
\bigskip
\caption{As in Fig.~\protect\ref{fig:showers_zenith}, but for tau
events with $E^{\rm thr}_{\tau} = 2.5\times 10^6~\gev$.
\label{fig:taus_zenith}
}
\end{figure}

The zenith angle distributions of SM and black hole events for
showers, muons, and taus are given in Figs.~\ref{fig:showers_zenith},
\ref{fig:muons_zenith}, and \ref{fig:taus_zenith}, respectively.
Black hole production makes the Earth even more opaque to ultra-high
energy neutrinos, reducing the up-going rate, while increasing the
probability of interaction, and thereby the event rate, of down-going
neutrinos. In all cases, up-going events rates are below SM
contributions. No sensitivity to black hole production is then
expected when looking for up-going events.

The characteristic features of the zenith angle distribution of
showers compared to the distributions of muons and taus are also
noteworthy.  For showers, vertical fluxes are less attenuated than
horizontal ones, and so the angular distribution is maximized for
vertically down-going events.  In contrast, as noted in
Sec.~\ref{sec:SM}, vertically down-going muon and tau events cannot
benefit from muon and tau ranges beyond the depth of the detector
($\sim 2$ km).  The optimal direction for muon and tau events is
therefore closer to the horizon.  In scenarios with enhanced cross
sections such as the one we are exploring in this paper, the neutrino
flux is attenuated even in the horizontal direction as can be seen in
Figs.~\ref{fig:muons_zenith} and \ref{fig:taus_zenith}.  As a
consequence, the quasi-horizontal direction with small but positive
$\cos\theta_z$, where the effects of attenuation are maximally offset
by lepton range, leads to the largest rates.  For muons, this optimal
zenith angle is around $\cos\theta_z \approx 0.2$.  For taus, with
even longer ranges, it is $\cos\theta_z \approx 0.4$.  These are
general considerations that apply to the detection of ultra-high
energy neutrinos in any scenario predicting enhanced cross sections
with respect to the SM cross sections.

\begin{figure}[tb]
\postscript{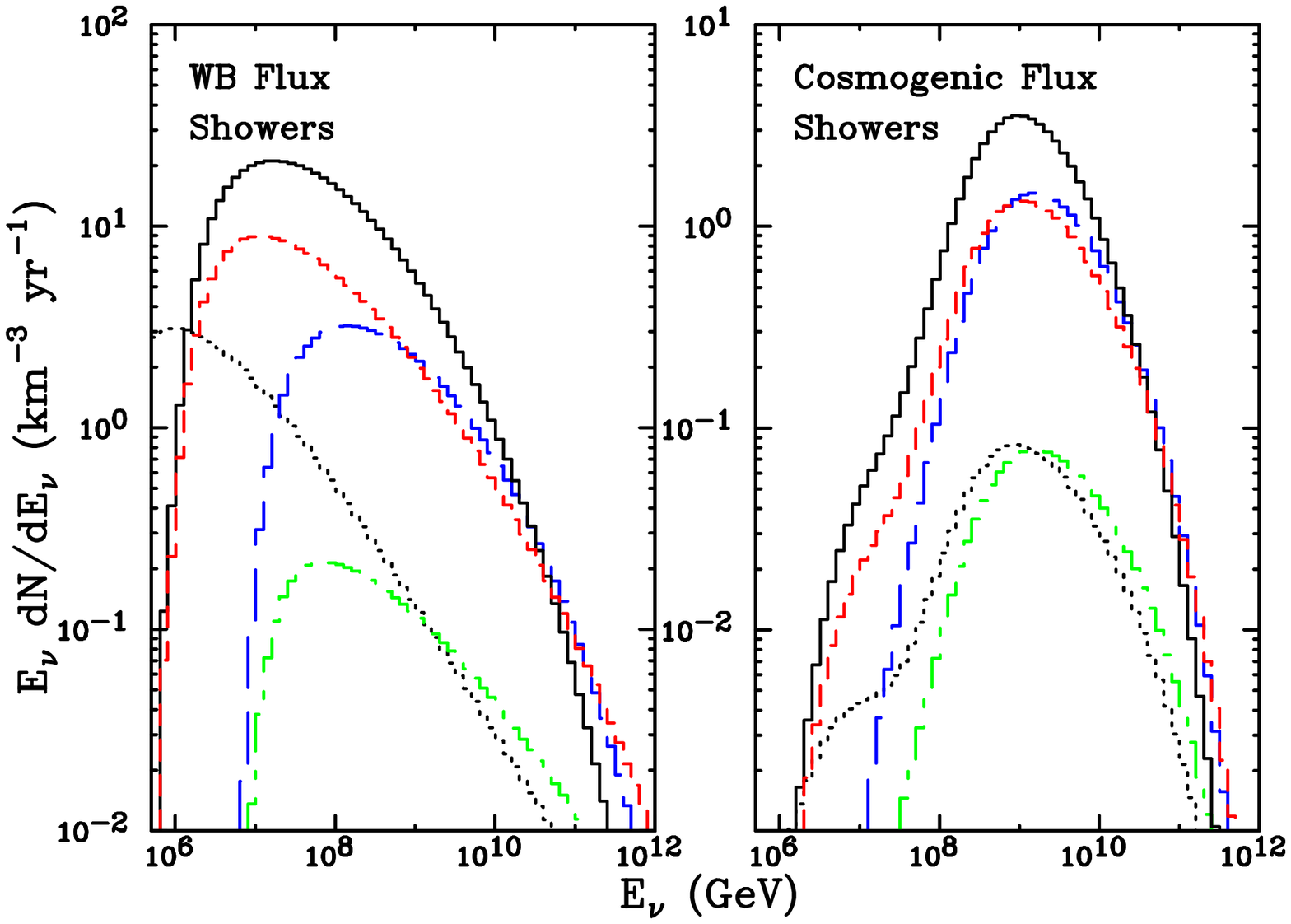}{0.67}
\caption{Energy distribution of down-going shower events for $\md =
1~\tev$, $E^{\rm thr}_{\rm sh} = 500~\tev$, and $(n, \xmin,
\hat{\sigma}) = (6,1,\pi r_s^2)$ (solid), $(6,3,\pi r_s^2)$ (long
dash), $(6,1,\pi r_s^2 e^{-I})$ (short dash), $(6,3,\pi r_s^2 e^{-I})$
(dot dash). Also shown is the standard model prediction (dotted).
\label{fig:showers_edistr_down}
}
\end{figure}

\begin{figure}[tb]
\postscript{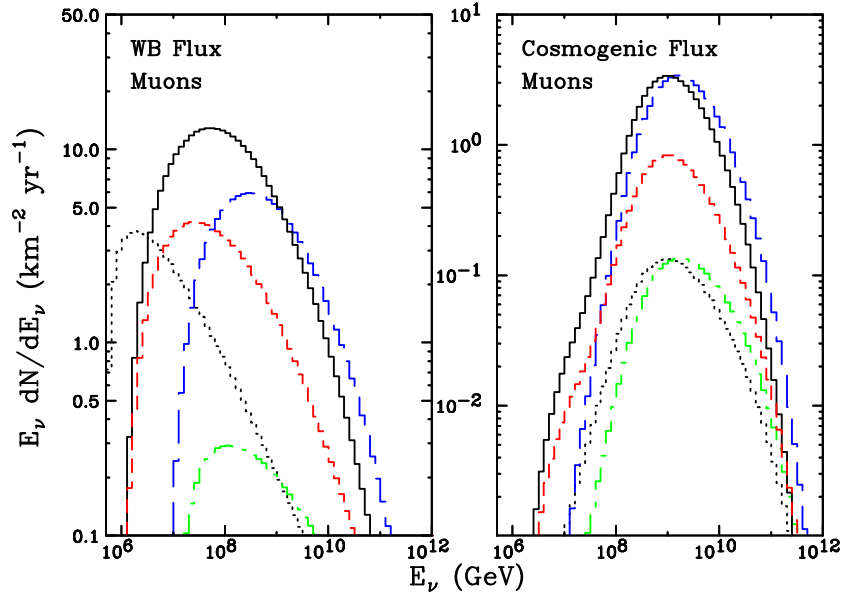}{0.67}
\caption{As in Fig.~\protect\ref{fig:showers_edistr_down}, but for
down-going muon events with $E^{\rm thr}_{\mu} = 500~\tev$.
\label{fig:muons_edistr_down}
}
\end{figure}

\begin{figure}[tb]
\postscript{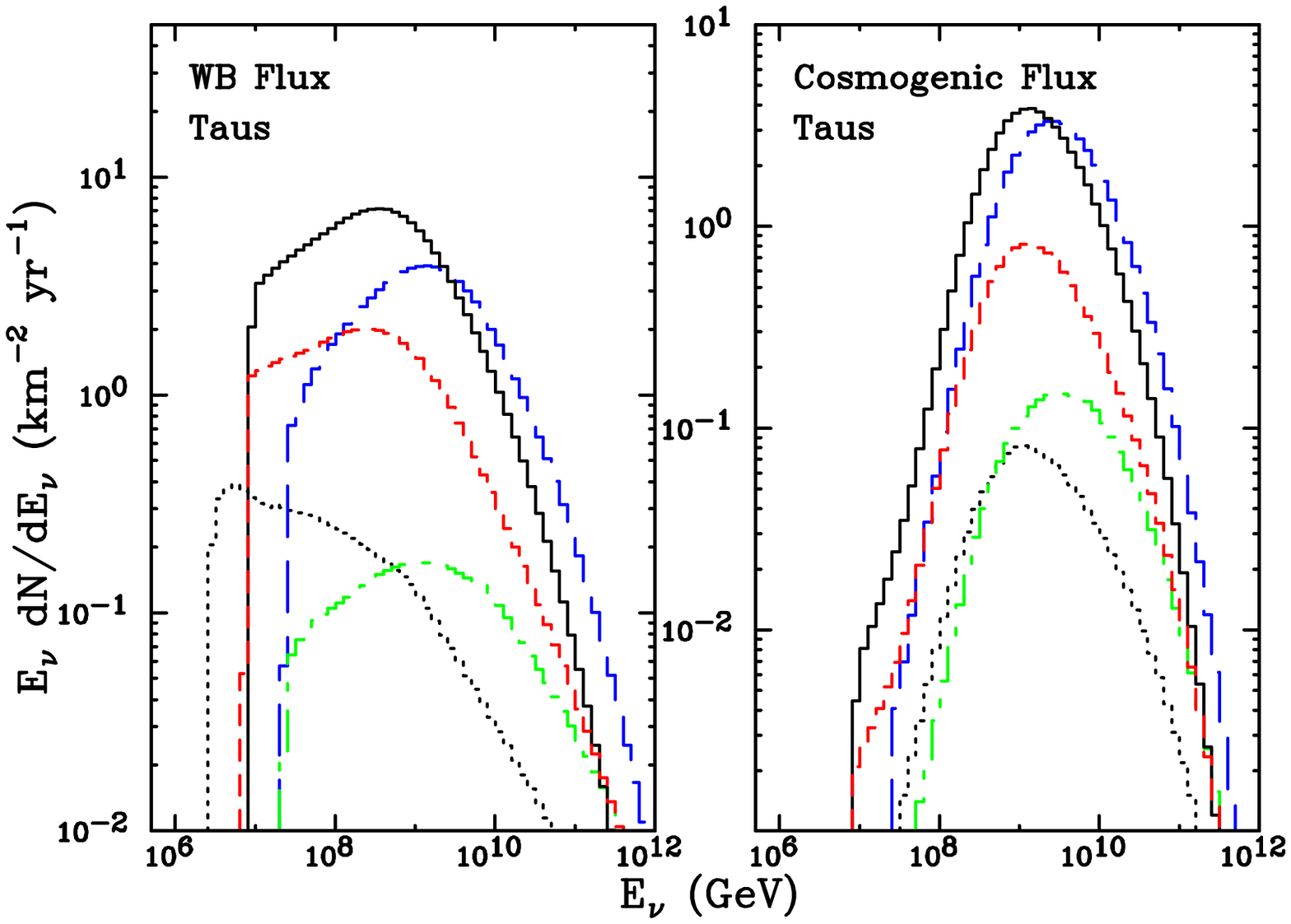}{0.67}
\caption{As in Fig.~\protect\ref{fig:showers_edistr_down}, but for
down-going tau events with $E^{\rm thr}_{\tau} = 2.5\times 10^6~\gev$.
\label{fig:taus_edistr_down}
}
\end{figure}

Energy distributions for the various channels are given in
Figs.~\ref{fig:showers_edistr_down}, \ref{fig:muons_edistr_down}, and
\ref{fig:taus_edistr_down}. The energy distributions are clearly
sensitive to $\xmin$ and the presence or absence of exponential
suppression in the parton cross section.  Different types of events
therefore probe different and complementary aspects of black hole
production and decay. By measuring the down-going energy distributions
with reasonable statistics, the IceCube detector may be able to
discriminate between the different possibilities.  Note that to
facilitate comparison with conventional presentations of SM rates at
neutrino telescopes, we have plotted distributions in initial neutrino
energy.  Shower energies are related to these neutrino energies in a
fairly direct way, and the shower energy distributions will have
roughly the same shape.  The energy distributions of through-going
leptons may be significantly distorted, however, as the lepton energy
is a function of black hole multiplicity as well as the distance the
lepton propagates before reaching the detector.  Note also that, while
shower energies should be well-measured, lepton energy measurements
present significant challenges, especially for taus.  Lollipop tau
events will be a clear tau signature with well-measured energy.
However, determining the energy of through-going tau events will be
difficult in IceCube as they typically do not radiate and, therefore,
appear similar to minimum ionizing muons with energy $\sim 200~\gev$.
More detailed detector simulations will be necessary to effectively
use the information from through-going taus described in this paper.

\section{Prospects and Conclusions}

In the presence of TeV-scale gravity and extra dimensions, ultra-high
energy cosmic neutrinos may produce microscopic black holes in the
surface of the Earth.  We have explored the prospects for detecting
such events in neutrino telescopes.  We considered contained showers,
through-going muons, and also through-going taus.

The rates for a $\km^3$ detector, such as IceCube, are of the order of
a few per year in each channel for $\md$ near current bounds.  These
rates are well above SM backgrounds, and provide a significant
opportunity to observe black hole production.  In several years of
data taking at a kilometer-scale neutrino telescope, it should be
possible to probe most of the cases described in this paper: values of
$M_D$ well above 1 TeV may be within reach, and observable event rates
are even possible for exponentially suppressed parton cross
sections. At the same time, these results imply that black hole
observation at ANTARES and NESTOR appears unlikely, given their
effective areas of $\alt 0.1~\km^2$.  The RICE
experiment~\cite{Kravchenko:2001id}, which aims to detect coherent
Cerenkov radiation from electromagnetic and hadronic showers at
frequencies of 100 MHz to 1 GHz, provides another interesting and
complementary probe with effective volume for shower detection
reaching $\sim 1~\km^3$ above a few PeV.

The expected rates for black hole events are, of course, highly
sensitive to the choice of neutrino flux. As this work was being
completed, a paper also discussing black holes at neutrino telescopes
appeared~\cite{Kowalski:2002gb}.  There, results for two fluxes are
presented: a ``limit from hidden sources'' flux and a cosmogenic flux.
The first flux is roughly 3 times higher than the limit placed by the
AMANDA experiment on diffuse fluxes~\cite{Andres:2001ty}.  The WB flux
we use is much more conservative and is roughly 2 orders of magnitude
smaller. A variety of cosmogenic
fluxes~\cite{Yoshida:pt,Protheroe:1995ft} are also considered in
Ref.~\cite{Kowalski:2002gb}.  Our ratios of black hole to SM rates are
in agreement, but here, too, it appears we have chosen a more
conservative representative flux, leading to a factor of 3 to 5 fewer
shower and muon events from cosmogenic neutrinos.  The possibility of
tau events, exponentially suppressed cross sections, and angle and
energy distributions were not addressed in
Ref.~\cite{Kowalski:2002gb}.

Relative to IceCube, the prospects for black hole detection appear to
be slightly brighter at the Auger Observatory, expected to be
completed by 2004, where tens of events per year may be
discovered~\cite{Feng:2001ib,Ringwald:2001vk,Anchordoqui:2001cg}.
IceCube is scheduled to reach its ultimate goal of $1~\km^3$ in 2009,
but it will collect data continuously as it grows in stages, beginning
with the existing AMANDA II detector, roughly of area $0.1~\km^2$, and
reaching a volume of around $0.5~\km^3$ in 2006.  It is difficult to
model this time-varying detection volume, but by 2007, an integrated
exposure of roughly $1~\km^3~\yr$ will have been achieved.  Given the
rates of Tables~\ref{table:I}-\ref{table:III}, we conclude that, for
scenarios with $\md$ near current bounds, IceCube may detect black
hole events along with the Auger Observatory before the LHC begins
operation.

Of course, enhanced event rates alone are not necessarily a signal of
new physics.  In particular, the uncertainty of astrophysical neutrino
fluxes makes it difficult to determine whether observed events result
from the SM or new physics based on counting experiments alone.
Comparisons of quasi-horizontal showers and Earth-skimming neutrino
rates provide a powerful discriminant in ground
arrays~\cite{Feng:2001ib,Anchordoqui:2001cg}.  However, the
possibility of measuring relative event rates in three separate
channels at IceCube provides a direct, unique, and complementary tool
for distinguishing black hole events from the SM and other new physics
possibilities.  As noted in Sec.~\ref{sec:distributions}, angular and
energy distributions will also be helpful to separate the signal of a
large astrophysical flux from the signal of new physics.  The
separation of these channels in ground arrays and fluorescence
detectors is extremely difficult.

Given sufficient statistics, then, the information provided by IceCube
may provide information on black hole branching ratios, and the
various angle and energy distributions may also help distinguish
various properties of black hole production and decay.  Black hole
production may be the best prospect for discovering extra dimensions
in the near future, and neutrino telescopes may be an excellent tool
for studying them.

\begin{acknowledgments}
We thank Gonzalo Parente for invaluable help with implementing the
CTEQ parton distribution functions in our simulation, Ralph Engel for
providing the cosmogenic neutrino flux, and Alfred Shapere for helpful
conversations.  J.A.-M. is supported by NASA Grant NAG5-7009.  The
work of J.L.F. was supported in part by the Department of Energy under
cooperative research agreement DF-FC02-94ER40818. This work was
supported in part by DOE grant No.~DE-FG02-95ER40896 and in part by
the Wisconsin Alumni Research Foundation.
\end{acknowledgments}



\end{document}